\documentclass[preprint2,twoside]{hwo}

\usepackage{lipsum}
\usepackage{chemformula}

\input{hwo.h}

\begin{document}

\title{\textbf{\LARGE Pollux test bench: from NUV to FUV spectropolarimetric measurements}}
\author {\textbf{\large Adrien Girardot,$^{1}$ Coralie Neiner,$^1$ Jean-Michel Reess,$^1$ Olivier Dupuis,$^1$, Margarita Carret,$^1$ }}
\affil{$^1$\small\it LIRA, Observatoire de Paris, Université PSL, Sorbonne Université, Université Paris Cité, CY Cergy Paris Université, CNRS,92190 Meudon, France}

\begin{abstract}
  Pollux is a high-resolution spectropolarimeter proposed by an European consortium for HWO. The current design of Pollux features four spectropolarimetric channels, three of which are in the UV range. For the near-UV (NUV) [236-472 nm] and mid-UV (MUV) [118-236 nm] channels, the polarimeters consist of waveplates and prisms made of MgF2, a birefringent material. However, no such birefringent material can be used for the far-UV (FUV) channel [100-123 nm]. Therefore, the polarimeter for this FUV channel is composed solely of mirrors in an innovative assembly. In this talk, we aim to detail the architecture of the test bench that will allow us to validate the performance of these different polarimeters, as part of the HWO GOMaP. Given that we are working in the vacuum ultraviolet (VUV) range, the test bench operates in a vacuum chamber in a clean room. We will discuss the adaptable architecture of the bench based on wavelength and the measurement methodology that we will implement to test if the polarimeters achieve the precision of $10^{-3}$ required for the Pollux instrument. With this test bench, we will successfully increase the Technology Readiness Level (TRL) of UV spectropolarimeters and, for the first time, develop a means to test FUV spectropolarimetry.
  \\
  \\
\end{abstract}

\vspace{2cm}

\section{Introduction}
Pollux is a high-resolution spectropolarimeter proposed by a European consortium as a potential instrument aboard the Habitable Worlds Observatory (HWO). HWO is a flagship mission currently under study, with several telescope architectures being investigated, featuring apertures between 6 and 8 meters, and a planned launch after 2040 \cite{10.1117/12.3030015}.

The current design of Pollux covers a broad spectral range, from 100 to 1,800 nm, with a spectral resolution up to $R = 100,000$ \cite{10.1117/12.3020175}. In particular, it includes three channels in the ultraviolet (UV), extending into the far-ultraviolet (FUV, 100–123 nm), a spectral region of strong astrophysical interest but technically challenging to access. Previous studies have already presented the polarimeter design for this channel \cite{girardot_design_2024, LeGalUnknownTitle2019}, highlighting innovative approaches required to reach the scientific goals of Pollux.

In this context, the development and validation of the Pollux polarimeters constitute a key step toward increasing the Technology Readiness Level (TRL) of UV spectropolarimetry. The present work focuses on the experimental test bench that will support this effort.

\section{Test bench concept}
The primary objective of the test bench is to quantify the polarimetric performance of the Pollux polarimeters with a target precision of \(10^{-3}\). To accommodate the different spectral regions of interest, the bench features two configurations: one optimised for the MUV and NUV bands (120-290 nm) and an adapted setup for the FUV range (100–120 nm).

\begin{figure}[ht]
    \includegraphics[width=\columnwidth]{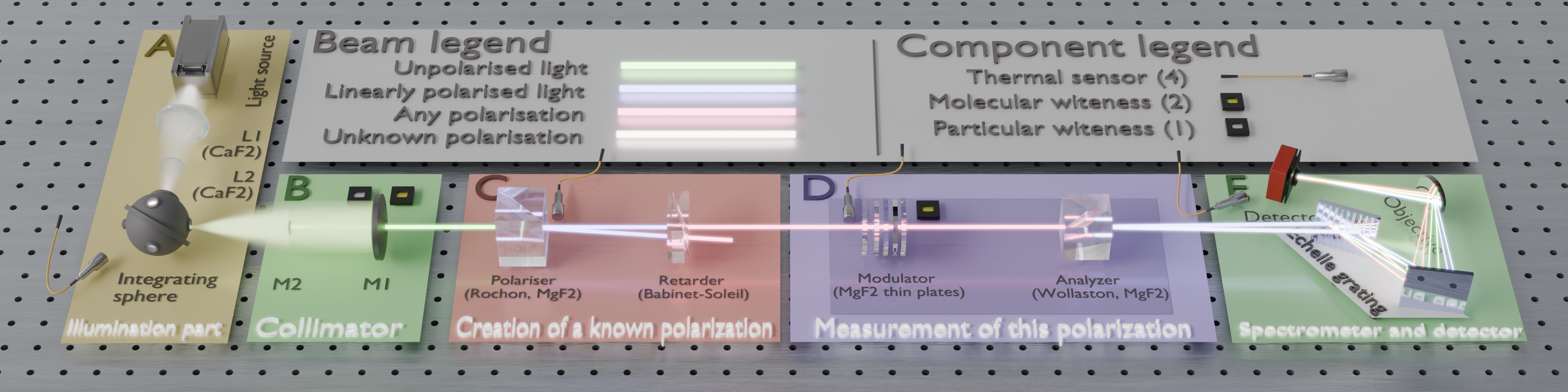}
    \caption{\small Concept of the MUV-NUV test bench
    \label{fig:shemabvanc}}
\end{figure}

\begin{figure}[ht]
    \includegraphics[width=\columnwidth]{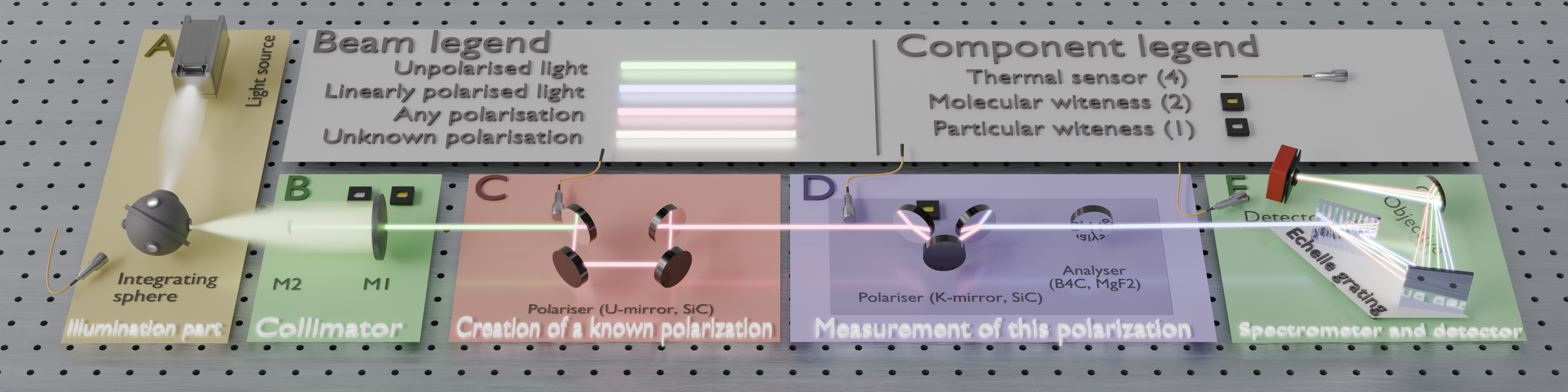}
    \caption{\small Concept of the FUV test bench
    \label{fig:schemabancfuv}}
\end{figure}

The experimental setup is composed of five blocks, represented Fig. \ref{fig:shemabvanc} and \ref{fig:schemabancfuv}. Blocks A and B generate a collimated unpolarized beam serving as the light input. Block C introduces any desired polarisation state to this beam via dedicated optics. The polarimeter under test is marked as block D and will measure the incident polarisation state with high accuracy. Finally, block E houses a spectrograph that disperses light for spectral and polarimetric analysis.

All optical components in this bench operate within a vacuum environment to prevent contamination and reduce absorption. The bench design accounts for the challenges of aligning and maintaining optical performance at these wavelengths, where reflective coatings and mirror alignment tolerances are critically demanding. 
\subsection{MUV-NUV bench}
The MUV-NUV test bench configuration is dedicated to characterising the CASSTOR polarimeter. CASSTOR is a nano-satellite that will be used as a demonstrator for Pollux UV polarimeters \cite{10.1117/12.3072789}. 

A deuterium lamp provides radiation with an initially undefined polarization state. The beam is injected, via $MgF_2$ and $CaF_2$ lenses, into an integrating sphere to ensure complete depolarization. A 1$\mu m$ point source is then collimated, producing a stable, depolarized collimated beam (end of Block B). Block C comprises a Wollaston prism and a Babinet–Soleil compensator, enabling the generation of any polarization states. The polarimeter under test, (Block D), employs $MgF_2$ waveplates and prisms, a well-known birefringent material transparent in the mid- and near-UV ranges. The optical path concludes with an echelle spectrograph with two gratings, an objective, and the detector.

It emits from 115 to 300 nm, enabling the testing of the CASSTOR polarimeter on its whole waveband and also covers the MUV range of Pollux. 

The optical path is carefully controlled to produce a highly collimated beam, ensuring uniform polarisation across the active polarimeter aperture. Initial alignments target minimising wavefront errors and beam divergence, critical factors influencing polarisation measurement accuracy. This configuration allows us to elevate the TRL of existing UV polarisation technologies by providing controlled, repeatable, and precise measurement conditions similar to those expected during operation on Pollux.

\subsection{FUV bench}
The FUV bench is specifically designed to address the absence of birefringent materials in the [100-118] nm spectral region and allowing characterisation of the mirror-based Pollux FUV polarimeter. The setup uses a deuterium plasma light source generated within the vacuum chamber to produce emission below 118 nm.

This test platform provides the first opportunity to experimentally validate the innovative FUV polarimeter architecture proposed for Pollux.

\section{Detector simulation}
The detector used in the test bench is a CIS120 sensor from Teledyne, selected for its sensitivity and suitability for UV applications. To ensure compatibility with the high vacuum environment, the detector’s printed circuit board (PCB) is coated with Mapsil, providing necessary vacuum compatibility.

\begin{figure}[ht]
    \includegraphics[width=\columnwidth]{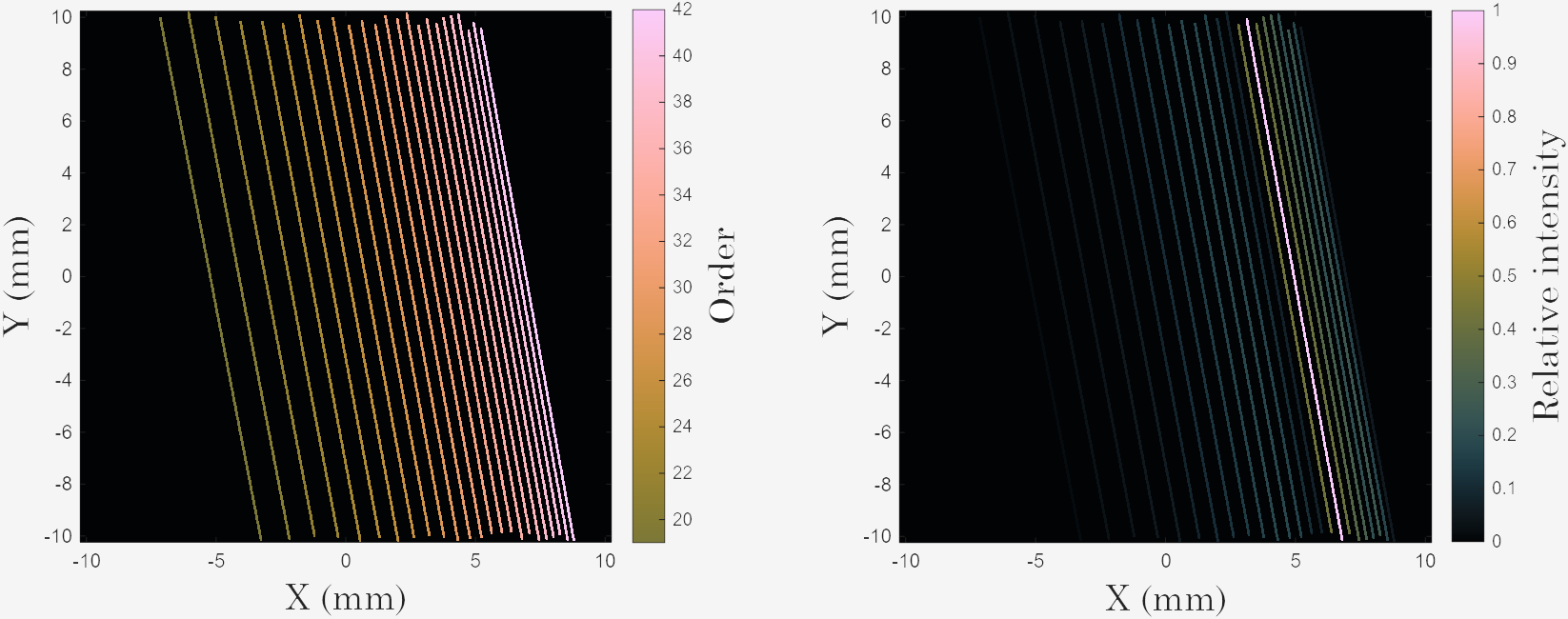}
    \caption{\small Detector simulation. \textbf{Left}: the colour represents the spectral order. \textbf{Right}: the colour represents the intensity.}
    \label{fig.simudet}
\end{figure}

Simulations of the detector output for the MUV-NUV bench configuration illustrate the spectral response expected during operation. On the left of Fig. \ref{fig.simudet}, each colour in the simulation corresponds to distinct spectral orders dispersed by the spectrograph. The right panel of Fig. \ref{fig.simudet} represents the signal distribution expected on the detector. 

\section{Operational test bench}
The entire test bench operates within a vacuum chamber to eliminate atmospheric absorption and contamination critical for ultraviolet measurements. The vacuum level of the chamber reaches approximately \(2 \times 10^{-6}\) mbar in 12 minutes after pumpdown, providing a clean and stable environment.
Contamination is regularly checked thanks to molecular and particular contamination witnesses. 
\begin{figure}[h]
    \centering
    \includegraphics[width=0.48\textwidth]{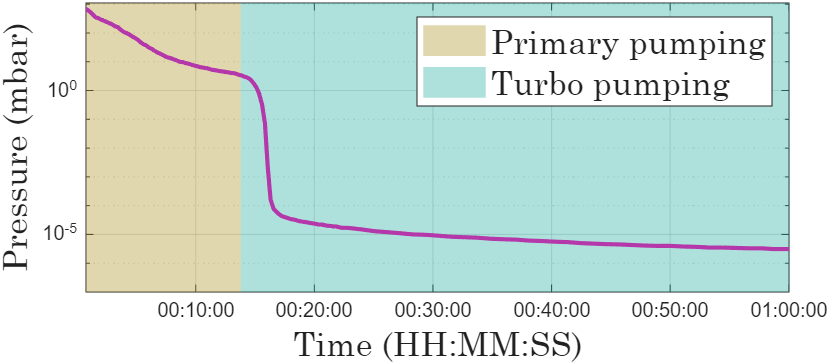}
    \caption{Pumpdown inside the vacuum chamber}
    \label{fig:placeholder}
\end{figure}

\begin{figure}[ht]
    \includegraphics[width=\columnwidth]{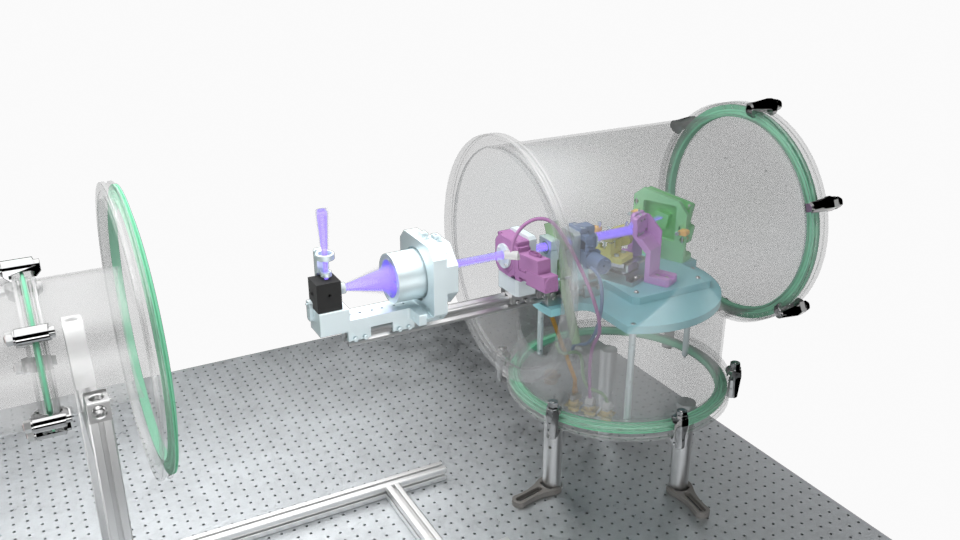}
    \caption{\small Test bench in the MUV-NUV configuration. Total length $\approx 1\;m$}
    \label{fig:cuve}
\end{figure}

Mechanical design facilitates accessibility: As shown in Fig. \ref{fig:cuve}, the chamber is mounted on rails, allowing access to internal optical elements. Components such as the collimator, polariser, Babinet-Soleil retarder, and analyser are pre-aligned externally before chamber closure. Fine alignment of the remaining optics is performed inside the vacuum chamber using nine vacuum-compatible piezoelectric motors, offering precise control over all necessary degrees of freedom.
The objective mirror, the most complex component in terms of alignment, is mounted on a five-axis stage complemented by an additional piezomotor for fine positional adjustments. The complete bench assembly spans approximately one meter in total length.

\begin{figure}[ht]
    \includegraphics[width=\columnwidth]{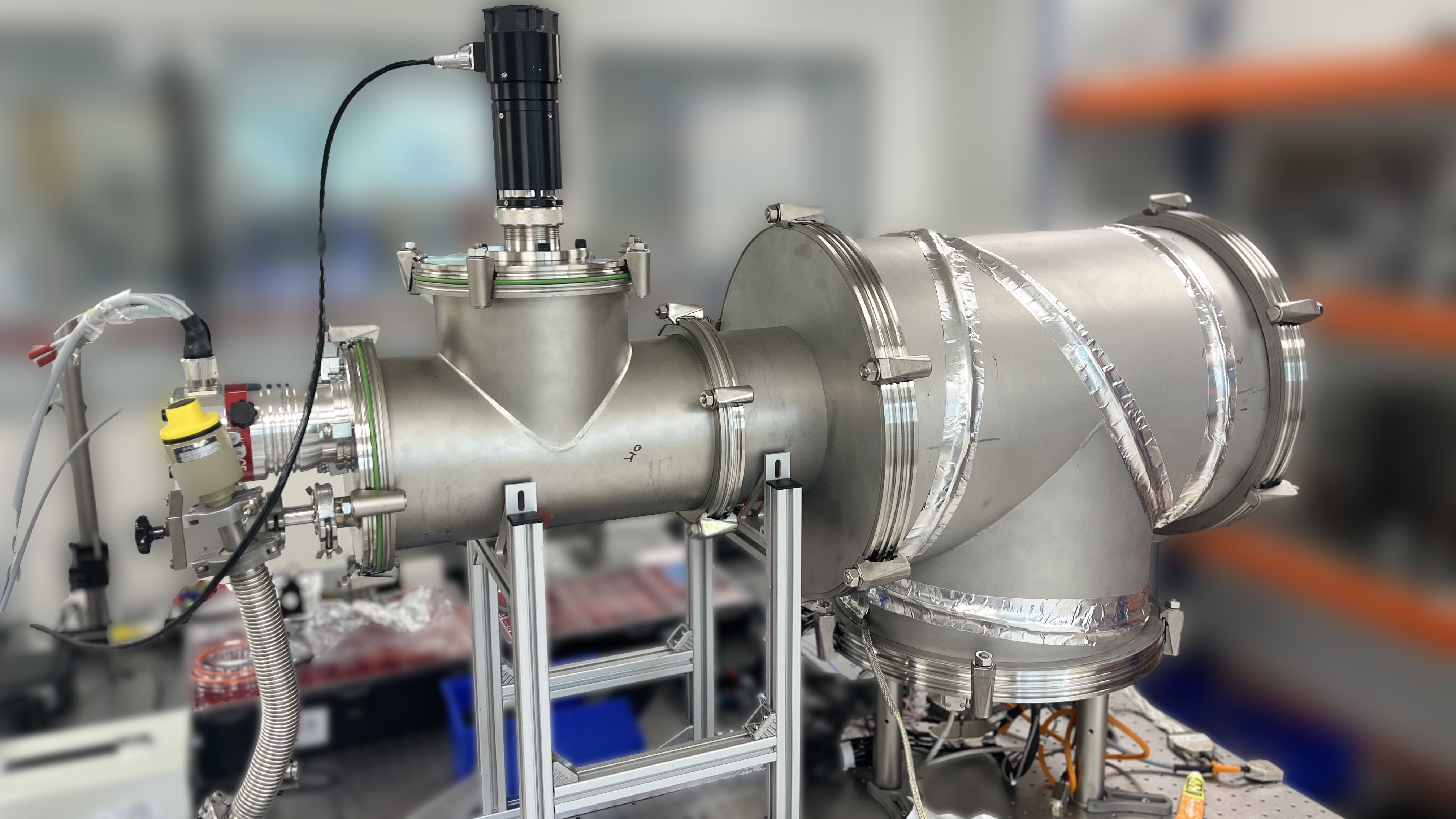}
    \caption{\small Vacuum chamber of the test bench
    \label{author_fig1_label}}
\end{figure}

Inside the vacuum chamber (artistic view on Fig. \ref{fig:cuve}), blocks A, B, C, and D are mounted on a rail, while block E is fixed on the blue platform.
        
\section{Example of subsystems alignments}
Prior to evaluating the polarimetric performance of the full system, precise alignment and calibration of individual optical subsystems are essential. This ensures that each component performs within the strict tolerances required to achieve the overall polarimetric precision of \(10^{-3}\). The following subsections illustrate key examples of such calibrations and alignments, focusing on critical elements including the Babinet-Soleil retarder and the collimator assembly. These subsystems form the core of the optical path of the test bench and directly affect the fidelity of polarisation generation and, thus, measurement.

\subsection{Calibration of the Babinet-soleil retarder}
The Babinet-Soleil retarder used in the test bench consists of three birefringent plates, with one plate mounted on a motorised stage to allow continuous tuning of the retardance. Calibration was performed at a wavelength of 632.8 nm using an external polarimetric setup, enabling precise determination of the zero-retardance position.

This calibration achieved a positioning accuracy of 0.01 mm, corresponding to a retardance thickness uncertainty of \(4.36 \times 10^{-5}\) mm. Achieving such fine calibration is vital for ensuring that the retarder can produce well-defined polarisation states, which are crucial calibration references for the polarimeters under test.

\begin{figure}[ht]
    \includegraphics[width=\columnwidth]{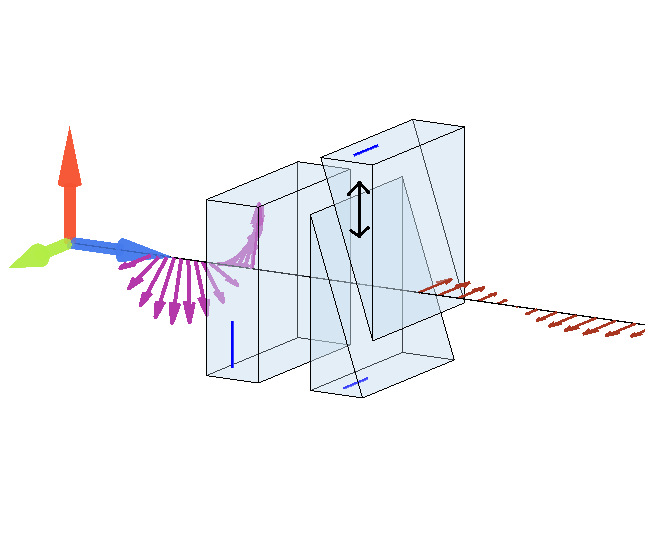}
    \caption{\small Babinet-soleil retarder principle
    \label{author_fig1_label}}
\end{figure}

\subsection{Collimator alignment}

\begin{figure}[ht]
    \includegraphics[width=\columnwidth]{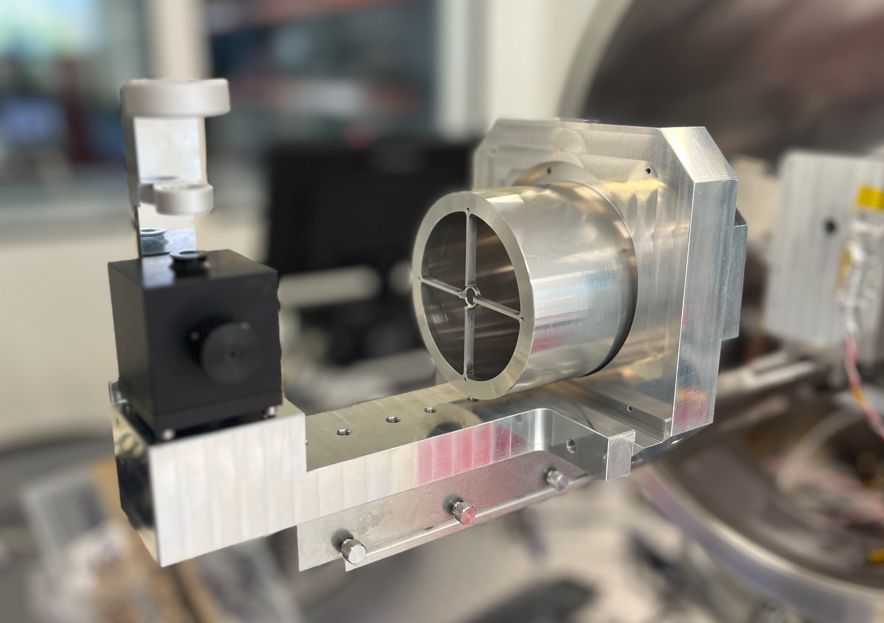}
    \caption{\small Collimator subsystem, the black cube is the integrating sphere. 
    \label{author_fig1_label}}
    \label{fig.collima}
\end{figure}

The collimator subsystem is composed of two mirrors, each with 2 degrees of freedom, enclosed in one single mechanical mount (see Fig. \ref{fig.collima}. Precise alignment is essential to produce a collimated beam with minimal wavefront aberrations.

Alignment is achieved using a QWLSI Phasics wavefront sensor operating at 632.8 nm, which facilitates wavefront error measurement with high sensitivity. Through iterative adjustments, the system reached an angular alignment precision of 30 arcseconds and a positional accuracy better than 1.5 $\mu m$. These tolerances ensure that the optical beam quality meets the stringent requirements for high-precision polarimetric measurements.

\section{Conclusion}

We have developed a vacuum ultraviolet test bench capable of characterising spectropolarimeters across a broad ultraviolet spectrum extending from the near-UV to the far-UV. Its design enables reconfiguration between the MUV-NUV bench, designed for classical birefringent polarimeters, and the FUV bench, dedicated to the novel mirror-based polarisation devices.

Early calibration and alignment efforts demonstrate the feasibility of achieving the polarimetric precision requirements necessary for Pollux. This infrastructure represents a critical advancement towards elevating the Technology Readiness Level of UV spectropolarimetric instrumentation and lays the groundwork for future in-orbit implementation on the Habitable Worlds Observatory mission.

\bigskip

{\bf Acknowledgements.} The authors thank CNES for their financial support.

\bibliography{author.bib}

@article{LeGalUnknownTitle2019,
    doi = {10.1117/12.2536094},
    year = {2019},
    month = {JUL},
    title = {Far ultra-violet polarimeter by reflection for Pollux (LUVOIR)},
    journal = {ICSO 2018},
    publisher = {SPIE},
    author = {Maelle LeGal and Coralie Neiner and Martin Pertenais.}
}

@inproceedings{girardot_design_2024,
	location = {Yokohama, Japan},
	title = {Design of a {FUV} polarimeter for Pollux aboard {HWO}},
	doi = {10.1117/12.3017994},
	pages = {147},
        year = {2024},
	booktitle = {Space Telescopes and Instrumentation 2024},
	publisher = {{SPIE}},
	author = {A. Girardot and Coralie Neiner and Jean-Michel Reess},
}

@inproceedings{10.1117/12.3020175,
author = {Eduard Muslimov and Coralie Neiner and Jean-Claude Bouret},
title = {{Optical design options for Pollux: UV spectropolarimeter project for the Habitable Worlds Observatory}},
volume = {13093},
booktitle = {Space Telescopes and Instrumentation 2024: Ultraviolet to Gamma Ray},
editor = {Jan-Willem A. den Herder and Shouleh Nikzad and Kazuhiro Nakazawa},
organization = {International Society for Optics and Photonics},
publisher = {SPIE},
pages = {130933D},
year = {2024},
doi = {10.1117/12.3020175},
URL = {https://doi.org/10.1117/12.3020175}
}

@inproceedings{10.1117/12.3030015,
author = {John Z. Lou and David C. Redding and Scott Basinger and Branden Dube and Jonathan Tesch and Andy Kee and Carl Nissly and Mitch Troy},
title = {{Habitable world observatory modeling optical and metrology models, and performance predictions}},
volume = {13129},
booktitle = {Optical Modeling and Performance Predictions XIV},
editor = {Mark A. Kahan and Catherine Merrill},
organization = {International Society for Optics and Photonics},
publisher = {SPIE},
pages = {131290K},
year = {2024},
doi = {10.1117/12.3030015},
URL = {https://doi.org/10.1117/12.3030015}
}

@inproceedings{10.1117/12.3072789,
author = {Coralie Neiner and Adrien Girardot and Jean-Michel Reess},
title = {{Space UV polarimeters}},
volume = {13699},
booktitle = {International Conference on Space Optics — ICSO 2024},
editor = {Fr{\'e}d{\'e}ric Bernard and Nikos Karafolas and Philippe Kubik and Kyriaki Minoglou},
organization = {International Society for Optics and Photonics},
publisher = {SPIE},
pages = {1369928},
year = {2025},
doi = {10.1117/12.3072789},
URL = {https://doi.org/10.1117/12.3072789}
}

\end{document}